# One-directional thermal transport in densely aligned single-wall carbon nanotube films

*Shingi Yamaguchi[1], Issei Tsunekawa[1], Natsumi Komatsu[2], Weilu Gao[2], Takuma Shiga[1], Takashi Kodama[1], Junichiro Kono[2,3,4,#], Junichiro Shiomi[1] ***

1. Department of Mechanical Engineering, The University of Tokyo, 7-3-1, Hongo, Bunkyo-ku, Tokyo 113-8656

2. Department of Electrical and Computer Engineering, Rice University, Houston, Texas 77005, U.S.A.

3. Department of Physics and Astronomy, Rice University, Houston, Texas 77005, U.S.A.

4. Department of Materials Science and NanoEngineering, Rice University, Houston, Texas 77005, U.S.A.

[#]kono@rice.edu, *shiomi@photon.t.u-tokyo.ac.jp

ABSTRACT: Individual carbon nanotubes (CNTs) possess extremely high thermal conductivities. However, the thermal conductivities and their anisotropy of macroscopic assemblies of CNTs have so far remained small. Here, we report results of directional thermal transport measurements on a nearly-perfectly aligned CNT film fabricated via controlled vacuum filtration. We found the thermal conductivity to be 43 ± 2.2 W m$^{-1}$ K$^{-1}$ with a record-high thermal anisotropy of 500. From the temperature dependence of the thermal conductivity and its agreement with the atomistic phonon transport calculation, we conclude that the effect of intertube thermal resistance on heat conduction in the alignment direction is negligible because of the large contact area between CNTs. These observations thus represent ideal unidirectional thermal transport, i.e., the thermal conductivity of the film is determined solely by that of individual CNTs.

Thermal management in electronics is becoming more and more important as the degree of device miniaturization has reached a truly nanometer scale where the level of power dissipation is also extreme. Therefore, thermally conducting electronic nanomaterials are strongly required for more efficient heat dissipation. Carbon nanotubes (CNTs) are one of the most promising candidates, since individual CNTs have exhibited thermal conductivities ($\kappa$) over $10^3$ W m$^{-1}$ K$^{-1}$ [1–5]. There have also been many thermal conductivity studies of CNT assemblies. In particular, $\kappa$ has been measured for aligned CNT samples prepared either by direct chemical vapor deposition growth[6-16] or post-processing of synthesized CNTs, such as mechanical processing[17–22], direct spinning[23], and magnetic alignment[24–26]. However, the $\kappa$ values reported for aligned CNT materials have so far been limited to tens to hundreds of W m$^{-1}$ K$^{-1}$, significantly lower than those for individual CNTs. This drastic difference has been attributed to structural issues such as low volume fractions (0.5 ~ 50 %), high defect densities, and low degrees of alignment, which make the intertube thermal resistance limit the film thermal conductivity. Therefore, it is essential to eliminate these structural deficiencies to utilize the high $\kappa$ of individual CNTs in aligned CNT assemblies.

In addition to the general need for materials with high thermal conductivity, there is also a specific demand for materials with anisotropic thermal conductivity that can direct heat flow only in a certain direction. For example, when spreading the heat from a chip on a circuit board, such directional heat flow can prevent heat-sensitive components from being damaged by excess heat conducted from heat-generating parts[27]. CNTs are clearly a good candidate because of their uniquely one-dimensional structure. However, unexpectedly, the thermal anisotropy of aligned CNT assemblies has not exceeded 100 in previous reports[6-9,16–20,24,25].

Here, we demonstrate a record-high value of thermal conductivity anisotropy in macroscopic films of aligned and packed CNTs prepared by the recently developed controlled vacuum filtration (CVF) method[28,29]. Using the T-type and time-domain thermoreflectance (TDTR) methods, we obtained a thermal anisotropy of 500 at room temperature (R.T.), which is the largest value obtained among all previously studied macroscopic CNT assemblies[6-9,16–20,24,25]. Although the $\kappa$ in the alignment direction ($\kappa_{||}$) is lower than the highest reported values[6–26], further theoretical analysis reveals that the $\kappa_{||}$ is determined only by the $\kappa$ of the constituent CNTs and is not limited by intertube thermal resistance.

The CNT used in this study was unsorted arc-discharge CNT purchased from Carbon Solutions, Inc., and the highly aligned CNT films (Film 1) were prepared by the CVF method[28]. In addition, a poorly aligned CNT film (Film 2) and a randomly aligned CNT film (Film 3) were also prepared for comparison. Film 2 was prepared by filtrating the CNT dispersion with additional 6 mM of NaCl at a filtration speed four times higher than that in the case of Film 1. These changes alter the interaction between the CNTs and the filter membrane[28] and reduce the degree of alignment. Film 3 was prepared by filtration of a CNT dispersion without any additives or filtration control. $\kappa_\parallel$ for each sample was measured by the T-type method[1,30] (Figure S1). The ~200-nm-thick Films 1 and 2 were supported by the polyethylene terephthalate (PET) substrates, and their $\kappa_\parallel$ were calculated by subtracting the thermal conductance of the PET substrate from the thermal conductance of the CNT/PET (CNT and PET). Film 3 was ~30 μm thick and measured as a self-standing film. The cross-plane $\kappa$ ($\kappa_\perp$) of Film 1 was measured using TDTR[31,32] after coating the surface with a ~100-nm-thick Al transducer film. The details of our TDTR setup are described elsewhere[33].

Top-view Scanning Electron Microscope (SEM) images of each sample are shown in Figure 1. Film 1 (Figure 1a) had uniformly aligned CNTs, as in previous reports[28,29,34–39]. The alignment of Film 2 (Figure 1b and 1c) is much weaker. The focused SEM image in Figure 1b shows that the CNTs are oriented rather randomly but there are parts with higher density due to the local moderate alignment that gives rise to the different color contrast. This is more evident in the broad view (Figure 1c) where white lines indicate partial ordering in Film 2. The morphology of Film 3 (Figure 1d) was clearly different from that of the other two aligned samples; most CNTs existed as large bundles with a diameter of ~100 nm, and they formed a sparse network structure. The alignment degrees of Films 1 and 2 were evaluated by measuring the reduced linear dichroism ($LD^r$) with a 660 nm laser beam[38]. $LD^r$ was 0.68 for Film 1 and 0.040 for Film 2, so the alignment degree of Film 2 was less than one tenth of that of Film 1 (Table 1). The non-zero $LD^r$ value of Film 2 is also consistent with the ordering of CNTs seen in Figure 1c.

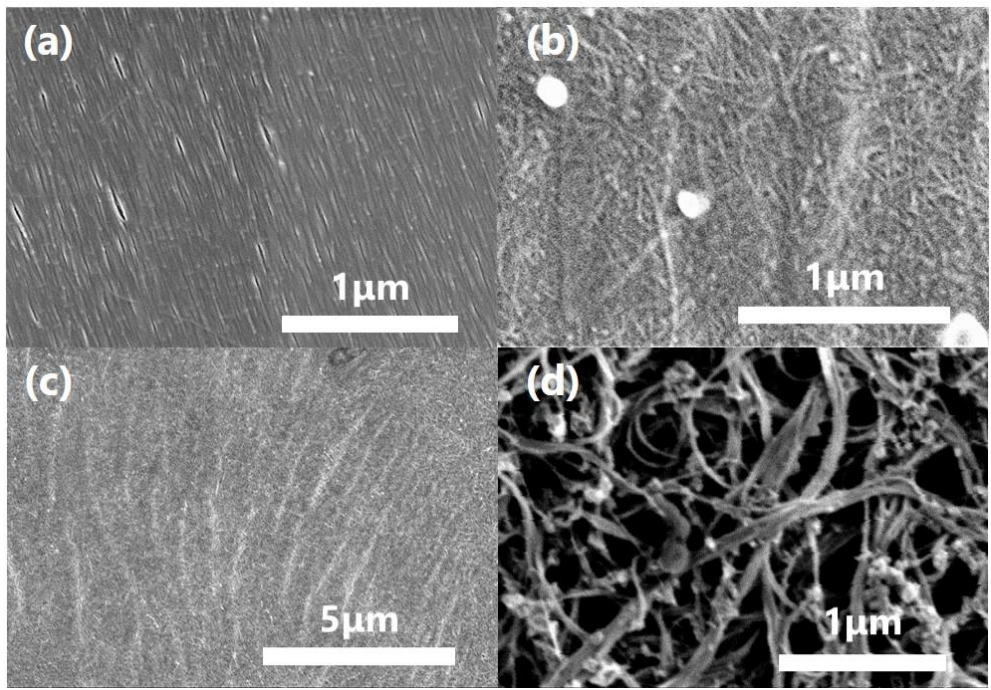

**Figure 1.** Scanning electron microscopy (SEM) images: (a) Film 1 (highly aligned film prepared by CVF method), (b) Film 2 (poorly aligned film prepared by CVF method with NaCl addition), (c) Film 2 with lower magnification than (b) and (d) Film 3 (randomly aligned film prepared by filtration of a CNT dispersion without any additives or filtration control.).

**Table 1.** $LD^r$ and $\kappa$ at R.T. of the three films studied.

|  | Film 1 (Highly aligned) | Film 2 (Poorly aligned) | Film 3 (Randomly aligned) |
|---|---|---|---|
| $LD^r$ | 0.68 | 0.040 | 0 |
| $\kappa_{\parallel}$ | 43 ± 2.2 W m$^{-1}$ K$^{-1}$ | 28 ± 1.2 W m$^{-1}$ K$^{-1}$ | 14 ± 2.8 W m$^{-1}$ K$^{-1}$ |
| $\kappa_{\perp}$ | 0.085 ± 0.017 W m$^{-1}$ K$^{-1}$ | - | - |

The $\kappa$ measurements of the three films were conducted over a temperature range from 50 to 300 K. First, $\kappa$ values of different samples at R.T. were compared to see any thermal property differences caused by the morphological differences, and are summarized in Table 1. The $\kappa$ of Film 1 in the alignment direction ($\kappa_{||,1}$ = 43 ± 2.2 W m$^{-1}$ K$^{-1}$) at R.T. was higher than those of Film 2 ($\kappa_{||,2}$ = 28 ± 1.2 W m$^{-1}$ K$^{-1}$) and Film 3 ($\kappa_{||,3}$ = 14 ± 2.8 W m$^{-1}$ K$^{-1}$). On the other hand, the $\kappa$ of Film 1 in the perpendicular direction ($\kappa_{\perp,1}$) was as low as 0.085 ± 0.017 W m$^{-1}$ K$^{-1}$, which is three orders of magnitude smaller than $\kappa_{||,1}$. This reveals that Film 1 had an extremely large thermal anisotropy ($\kappa_{||,1}/\kappa_{\perp,1}$) of 500, which is the largest reported value among aligned CNT films[6-9,16–20,24,25]. The highest and second-highest value of $\kappa_{||,1}$ and $\kappa_{||,2}$ show the $\kappa_{||}$ improvement of the film by the constituent CNT alignment, and the details of which will be discussed later. While the randamly aligned Film 3 showed lowest $\kappa_{||}$, its structure seen in Fig.1d contains seemingly aggregated CNT chunks. Therefore, it is difficult to separate the effect of aggregation from that of alignment. Therefore, the case of Film 3 is shown only to compare the absolute thermal conductivity value of conventional CNT mat with those of the aligned ones, and thus, the detailed heat conduction mechanism in Film 3 will not be discussed in this paper.

It is known that the $\kappa$ of CNT bundles are lower than those of individual CNTs due to the quenching of low-frequency phonon modes and small thermal conductance between CNTs[5,19,41,42]. However, despite the nearly perfect CNT alignment in Film 1, its $\kappa_{||,1}$ value at R.T. is still one order of magnitude smaller than the reported $\kappa$ of a single CNT bundle ($\kappa_{||,\text{bundle}}$)[41–43]. To understand the reason, the temperature dependence of $\kappa_{||,1}$ was examined in detail. First, the temperature dependence of $\kappa_{||,1}$ was compared with that of $\kappa_{||,\text{bundle}}$ measured in Ref. 43. As shown in Figure 2, the profiles normalized by the R.T. value show good agreement. In Ref. 43, CNT-CNT contact thermal resistance had negligible effects on $\kappa_{\parallel,\text{bundle}}$ because both the bundle itself and the constituent CNTs were 1 μm long; namely, all the CNTs seamlessly connected the two thermostats that suspended the bundle. Therefore, considering that the internal thermal conductance grows linearly with temperature[4] and the intertube thermal conductance depends weakly on temperature[40,43] in the range from 150 to 300 K, the agreement in temperature dependence between $\kappa_{\parallel,1}$ and $\kappa_{\parallel,\text{bundle}}$ suggests that the intertube thermal resistance has a limited effect on $\kappa_{\parallel,1}$.

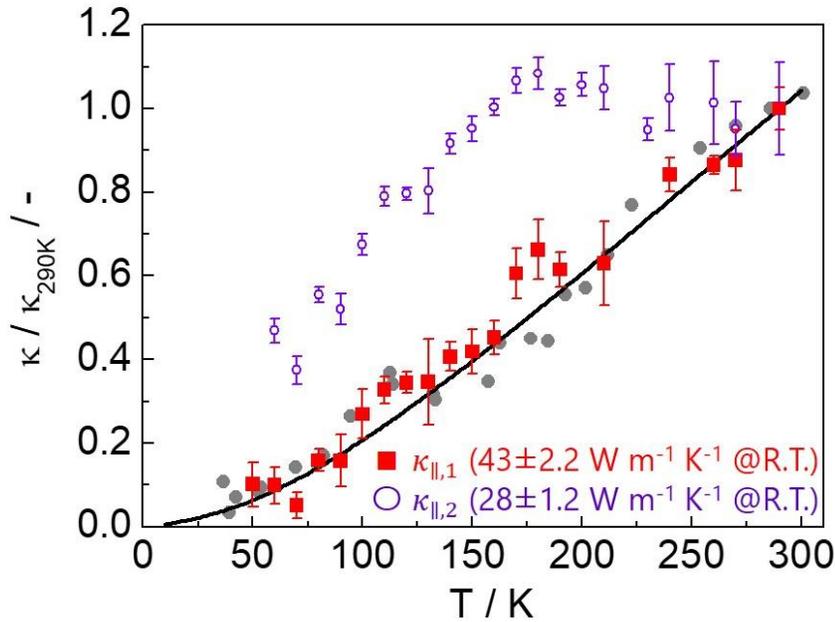

**Figure 2.** Temperature dependence of thermal conductivity. ■: Film 1 (alignment direction), ●: single CNT bundle[43], ○: Film 2 (alignment direction), solid line: simulated effective thermal conductivity of highly aligned CNT film. The values of thermal conductivity are normalized by the R.T. value for each material.

**Table 2.** List of the variables used in the calculation and their details.

| Variable | Explanation | Source |
|---|---|---|
| $G_\parallel$ | Internal thermal conductance of a CNT in the axial direction | AGF calculation |
| $G_\perp$ | Internal thermal conductance of a CNT in the perpendicular direction | - |
| $g$ | Intertube thermal conductance at an aligned CNT-CNT contact | AGF calculation |
| $g_{exp}$ | Actual value of the intertube thermal conductance at R.T. | Model in Fig.4b |
| $g'$ | Intertube thermal conductance at a CNT-CNT cross contact | Ref. 40,56 |
| $\kappa_{\parallel,1,eff}$ | The effective value of $\kappa_{\parallel,1}$ | Model in Fig.4a |
| $\kappa_{\parallel,2,eff}$ | The effective value of $\kappa_{\parallel,2}$ | Model in Fig.4c |
| $\kappa_{ind}$ | $\kappa$ of each CNT constituting the Film 1 | - |
| $\kappa_{\parallel,bundle}$ | $\kappa$ of CNT bundle | Ref. 43 |

To further verify this suggestion, the internal thermal conductance of a CNT in the axial direction ($G_\parallel$) and intertube thermal conductance at an aligned CNT-CNT contact ($g$) were calculated using the atomistic Green's function (AGF) method[45]. The variables used for the models and calculations below are summarized in Table 2. Note that since the CNT length is shorter than the average phonon mean free path[46–48], the phonon transport can be considered ballistic. As the average diameter ($d$) of consisting CNT of Film 1 is ~1.4 nm, a hexagonal unit cell consisting of (10,10) single-wall CNTs ($d$=1.36 nm) was used as a representative atomic scale model for the calculations. While Film 1 contains both metallic and semiconducting CNTs, the calculation result with metallic (10,10) CNTs is valid for the comparison because heat transport in CNTs is dominated by phonons than by electrons[49,50]. As shown in Figure S2a and S2b, the prepared cells for the alignment and perpendicular directions, respectively, were repeated twice in the section between the two leads. Periodic boundary conditions were applied in the directions of the cross section (Figure S3c and S3d). The interatomic interactions within and between CNTs were modeled by the Tersoff[51] and Lennard-Jones potentials[52], respectively; the potential parameters are described in the references.

The effective value of $\kappa_{\parallel,1}$ ($\kappa_{\parallel,1,\text{eff}}$) was estimated by the following equation based on a model shown in Figure 4a:

$$\frac{L}{\kappa_{\parallel,1,\text{eff}}} = \frac{1}{G_\parallel} + \frac{1}{g}\frac{A}{S} \quad \cdots(1)$$

Here, $L$, $A$, and $S$ represent the length, bottom area, and side area of a constituent CNT when CNTs are viewed as a honeycomb structure. The intertube term ($\frac{1}{g}\frac{A}{S}$) includes the area ratio $\frac{A}{S}$ as $G_\parallel$ and $g$ both correspond to the thermal conductance per unit area while the cross-sectional areas of heat conduction through CNTs and between CNTs are different. The temperature dependence of the calculated $\kappa_{\parallel,1,\text{eff}}$ is plotted in Figure 2, which agrees well with the experimental results. Here, the intertube term is around three orders of magnitude smaller than the internal term ($\frac{1}{G_\parallel}$), indicating that the temperature dependence of $\kappa_{\parallel,1,\text{eff}}$ reflects only that of internal thermal conductance. Therefore, from Eq. (1), $\kappa_{\parallel,1,\text{eff}}$ can be approximated as

$$\kappa_{\parallel,1,\text{eff}} = LG_\parallel \quad \cdots(2).$$

This equation can be further transformed to

$$\kappa_{\parallel,1,\text{eff}} = A\kappa_{\text{ind}} \quad \cdots(3)$$

where $k_{ind}$ represents $\kappa$ of each CNT constituting the Film 1. This shows that, when phonon transport is ballistic (i.e., $G_\parallel$ is constant), $\kappa_{\parallel,1,eff}$ is determined only by the length of the constituent CNTs, or in other words, only by $k_{ind}$. This explains the small observed $\kappa_{\parallel,1}$ (=43 W m$^{-1}$ K$^{-1}$); it is merely because the $k_{ind}$ of constituent CNTs are small due to their shorter length (around 200 nm) than those in the bundle in Ref. 43 (around 1 µm, giving $\kappa_{\parallel,bundle}$ = 200.2 W m$^{-1}$ K$^{-1}$), not because of the intertube thermal resistance. Note that it makes sense that the five-fold difference in the length results in the five-fold difference in $\kappa$ as thermal conductivity increases linearly with the length when heat conduction is ballistic[53].

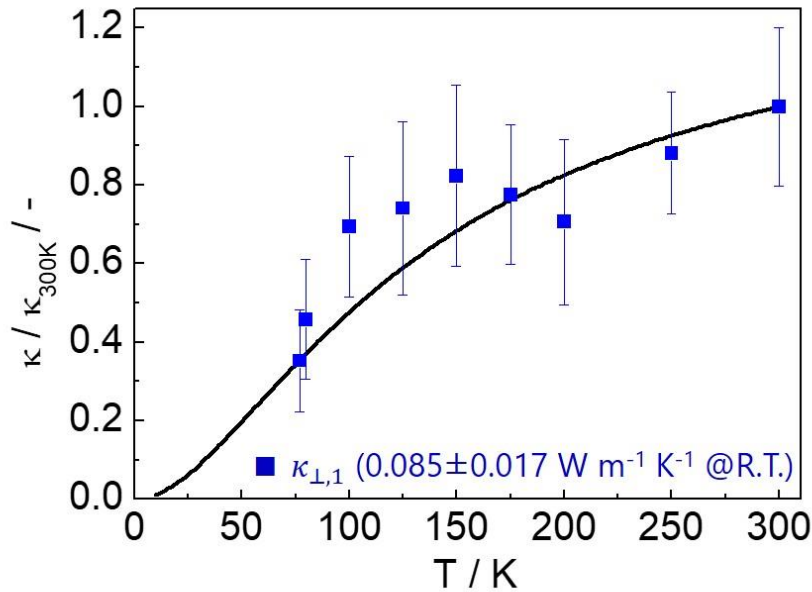

**Figure 3.** ■: Normalized thermal conductivity of Film 1 in the perpendicular direction in the temperature range of 77 to 300 K, solid line: normalized simulated thermal conductance of highly aligned CNT in the perpendicular direction.

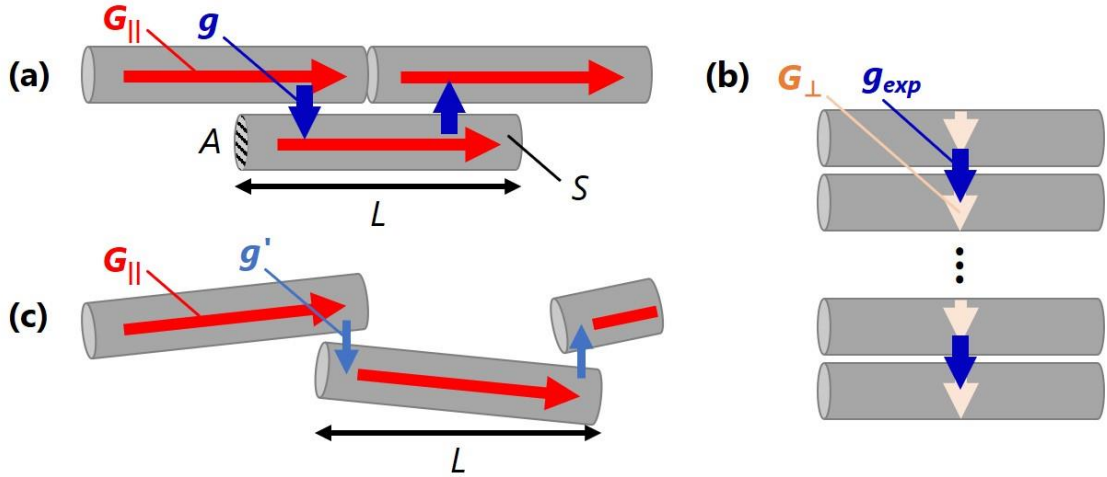

**Figure 4.** (a) Simulation models to estimate the effective thermal conductivity of Film 1 in the alignment direction. *L, A* and *S* represent length, bottom area and side area of a constituent CNTs. $G_\parallel$ represents thermal conductance of internal CNT and *g* and *g'* represent intertube thermal conductance at parallel/cross contact, respectively. (b) Simulation model for the calculation of intertube thermal resistance at CNT-CNT contact. (c) Simulation models to estimate the effective thermal conductivity of Film 2.

On the other hand, the intertube thermal resistance dominates the thermal conductivity of Film 1 in the cross-plane direction ($\kappa_{\perp,1}$). This can be confirmed by the good agreement in the temperature dependences of experimentally measured $\kappa_{\perp,1}$ and calculated *g* (Figure 3). Here, a quadratic increase at low temperatures and weak dependence at higher temperatures are observed, which is consistent with the temperature dependence of experimentally observed intertube thermal conductance between two CNTs[44]. Furthermore, the actual value of the intertube thermal conductance ($g_{exp}$) at R.T. was estimated from the experimental result ($\kappa_{\perp,1}$) using a simple model shown in Figure 4b. In the calculation, the internal thermal conductance of a CNT in the perpendicular direction ($G_\perp$) was assumed to be much greater than $g_{exp}$. Based on the model, $g_{exp}$ was calculated to be $1.5 \times 10^{-8}$ m² K W⁻¹, which is close to $g$ ($1.1 \times 10^{-8}$ m² K W⁻¹) and the other reported values calculated by molecular dynamics (MD) simulations[54–56]. This confirms that $\kappa_{\perp,1}$ is mainly determined by the intertube thermal resistance.

It is also worth noting that the temperature dependence of $\kappa_{\parallel,2}$ is different from that of $\kappa_{\parallel,1}$ (Figure 2). $\kappa_{\parallel,2}$ increases sharply in the low temperature regime, while it becomes almost constant above 180 K. It is tempting here to discuss the result in terms of the peak temperature shift as the peak temperature of CNT materials is known to appear between 300 to 400 K, resulting from the competition between the increase in heat capacity and decrease in the phonon mean free paths due to Umklapp scattering with increasing temperature. However, as the constituent CNT material of Film 2 is exactly the same as that of Film 1, it is unlikely that their Umklapp scattering rates or the heat capacities significantly differ.

Instead, this temperature dependence of $\kappa_{\parallel,2}$ can be explained by the stronger role of intertube thermal resistance at CNT-CNT cross contacts in Film 2 as shown in the simple model (Figure 4c). With this model, the effective $\kappa_\parallel$ of Film 2 ($\kappa_{\parallel,2,\text{eff}}$) can be estimated by the following equation:

$$\frac{L}{\kappa_{\parallel,2,\text{eff}}} = \frac{1}{G_\parallel} + \frac{1}{g'} \quad \cdots(4).$$

Here, $g'$ represents intertube thermal conductance at a CNT-CNT cross contact. Since the contact area is as small as the bottom area of CNT, the area ratio $\frac{A}{S}$ is not included in Eq. (4). According to the several reports clarifying the actual value of $g'$ from simulations, it generally ranges from $10^8$ to $10^9$ W K$^{-1}$ m$^{-2}$ [40,56], which is one to two orders of magnitude smaller than $G_{||}$ ($10^{10}$ W K$^{-1}$ m$^{-2}$ @ R.T.). Therefore, in contrast to the case of $\kappa_{||,1,\text{eff}}$, the intertube term $\frac{1}{g'}$ is dominant in Eq. (4), and the temperature dependence of $\kappa_{||,2,\text{eff}}$ is expected to follow that of $g'$. The temperature dependence of $g'$ has been investigated by both simulations[57] and experiments[44], where $g'$ showed a quadratic increase at low temperatures while the temperature dependence became very weak above 150 K, which is consistent with the behavior of $\kappa_{||,2}$ shown in Figure 2. This indicates that the difference in temperature dependence between $\kappa_{||,1}$ and $\kappa_{||,2}$ arises from the difference in leading mechanism of thermal resistance.

In conclusion, the $\kappa$ of a highly aligned CNT film synthesized by the CVF method was measured both in the alignment direction and perpendicular direction. The $\kappa$ value at R.T. was 43 ± 2.2 W m$^{-1}$ K$^{-1}$ and 0.085 ± 0.017 W m$^{-1}$ K$^{-1}$ in the alignment and perpendicular directions, respectively, yielding a thermal anisotropy of 500, the highest ever reported. Further analysis of the temperature dependence of $\kappa_{||}$ revealed that the effect of intertube thermal resistance, which is known to be large in pervious CNT films with weaker alignment, has a negligible influence on the $\kappa_{||}$ owing to the large intertube contact area realized by the nearly-perfect alignment, and $\kappa_{||}$ is determined only by $\kappa$ of the constituent CNT length. This also suggests that the $\kappa_{||}$ can be even greater with longer constituent CNTs.

SUPPLEMENTARY MATERIAL

See supplementary material for additional information regarding the synthesis method for CNT films, experimental setup, theoretical equation for thermal measurement, details for AGF calculation and absolute values of experiment/simulation data.

ACKNOWLEDGMENT

This research was supported by JSPS KAKENHI Grant Numbers 19H00744. N.K., W.G., and J.K. acknowledge support by the Basic Energy Science (BES) program of the U.S. Department of Energy through Grant No. DE-FG02-06ER46308 (for preparation of aligned carbon nanotube films), the U.S. National Science Foundation through Grant No. ECCS-1708315 (for optical measurements), and the Robert A. Welch Foundation through Grant No. C-1509 (for structural characterization measurements). We thank M. Ouchi (Iwase Group, Waseda University) for providing technical support with the processing of measurement sample with laser plotter and G. Timothy Noe II and Kevin Tian (Rice University) for proofreading the manuscript.

SUPPLEMENTARY MATERIAL

# One-directional thermal transport in densely aligned single-wall carbon nanotube films

*Shingi Yamaguchi[1], Issei Tsunekawa[1], Natsumi Komatsu[2], Weilu Gao[2], Takuma Shiga[1], Takashi Kodama[1], Junichiro Kono[2,3,4,#], Junichiro Shiomi[1]\**

1. Department of Mechanical Engineering, The University of Tokyo, 7-3-1, Hongo, Bunkyo-ku, Tokyo 113-8656

2. Department of Electrical and Computer Engineering, Rice University, Houston, Texas 77005, U.S.A.

3. Department of Physics and Astronomy, Rice University, Houston, Texas 77005, U.S.A.

4. Department of Materials Science and NanoEngineering, Rice University, Houston, Texas 77005, U.S.A.

**Methods**

*Materials*:

The highly aligned CNT films (Film 1) used in this study were prepared by the CVF method[1]. Arc-discharge CNTs purchased from Carbon Solutions, Inc. (P2-SWNT), were dispersed in surfactant solution. The dispersion was then vacuum filtered while the filtration speed was carefully controlled. In addition, a poorly aligned CNT film (Film 2) and a randomly aligned CNT film (Film 3) were also prepared for comparison. Film 2 was prepared by filtrating the CNT dispersion with additional 6 mM of NaCl at a filtration speed four times higher than that in the case of Film 1. Film 3 was prepared by filtration of a P2-SWNT dispersion without any additives or filtration control. Films 1 and 2 were transferred to either polyethylene terephthalate (PET) or Si substrates for thermal measurements by the T-type method or TDTR, respectively. The sample thickness was controlled to be ~200 nm on PET substrates and ~40 nm on Si substrates. Film 3 was ~30 µm thick and measured as a self-standing film. The alignment degrees of Films 1 and 2 were evaluated by measuring the reduced linear dichroism $LD^r = 2(A_\parallel - A_\perp)/(A_\parallel + A_\perp)^2$, where $A_\parallel$ and $A_\perp$ are the parallel and perpendicular absorbance, respectively. The average length of the P2-SWNT was statistically calculated based on the AFM images of the dispersed CNTs[1]. Its standard deviation was calculated to be 184 nm.

*Thermal measurements:*

In-plane $\kappa$ for each sample was measured by the T-type method[3,4]. Figure S1 shows our experimental setup. A platinum wire with a diameter of 10 µm was suspended between two electrodes (copper blocks), and the electrical potential between them was measured while a constant current was applied. The sample was suspended between the center of the wire and the heat sink (another copper block), and the measured values of potential with and without the sample were fitted with an analytical expression derived from our theoretical model to extract the in-plane $\kappa$ of the sample. The analytical expression was obtained by solving the heat conduction equation

$$\kappa_m A_m \frac{d^2 T_\pm(x)}{dx^2} + p'[1 + \alpha(T_\pm(x) - T_0)] = 0 \quad \cdots (S1)$$

$$(p' = \frac{I^2 R_0}{2L_m})$$

where $\kappa_m, A_m, L_m, \alpha$ are the thermal conductivity, cross-sectional area, half length, and temperature coefficient of resistance (TCR) of the Pt wire, and $I, T_0, T_\pm(x)$ are the electrical current, surrounding temperature and temperature of the wire at position $x$ (the signs corresponding to the coordinate when the origin is at the center of the wire), and $R_0$ is the electrical resistance of the wire when $T_\pm(x) = T_0$. The boundary conditions we impose are

(i) $T(x) = T_s \quad (-d \leq x \leq d) \quad \cdots (S2)$

(ii) $T(\pm L_m) = T_0 \quad \cdots (S3)$

(iii) $\kappa_s \frac{A_s}{L_s}(T_s - T_0) = \kappa_m A_m (\frac{dT_+}{dx}\big|_{x=d} - \frac{dT_-}{dx}\big|_{x=-d}) + \int_{-d}^{d} p'[1 + \alpha(T_+(x) - T_0)] dx \quad \cdots (S4)$

where $\kappa_s, A_s, L_s, d$ are the thermal conductivity, cross-sectional area, length, and half width of the sample, and $T_s$ is the temperature of sample at the contact with the wire. From the solution of Eq. (S1), we can obtain the average temperature of the wire $\overline{T(x)}$. Then the average electrical resistance $\overline{R}$ can be calculated using the TCR value α as

$$\overline{R} = R_0(1 + \alpha(\overline{T(x)} - T_0))$$

$$= \frac{R_0}{\sqrt{mL_m}} \frac{\frac{n(1+q)}{2\sqrt{mL_s}}\{1-\cos(\lambda-\delta)\} + \sin(\lambda-\delta) + \frac{nd}{2L_s}(\sin(\lambda-\delta) + \frac{2\sqrt{mL_s}}{n})}{\frac{nq}{2\sqrt{mL_s}}\sin(\lambda-\delta) + \cos(\lambda-\delta)} \quad \cdots (S5)$$

$$(m = \frac{\alpha p'}{\kappa_m A_m}, \quad n = \frac{\kappa_s A_s}{\kappa_m A_m}, \quad q = 1 - \frac{2dL_s \alpha p'}{\kappa_s A_s}, \quad \lambda = \sqrt{m}L_m, \quad \delta = \sqrt{m}d).$$

The thermal measurements were conducted over the temperature range from 50 to 300 K. While there was a temperature distribution along the wire, the positional dependence of $\kappa_m$ was ignored since $\kappa_m$ had little temperature dependence in the measured temperature range (i.e., 50 – 300 K, see Figure S3).

The measurement of Film 3 was conducted by suspending the film directly. Films 1 and 2 needed to be supported by a substrate, for which we chose PET because of its low thermal conductivity. For Films 1 and 2, after measuring the in-plane $\kappa$ of the CNT/PET (CNT and PET) sample, the thermal conductance of the CNT film ($K_{CNT}$) was obtained by subtracting the thermal conductance of the PET substrate from the thermal conductance of the CNT/PET $\left( K_{CNT} = K_{CNT/PET} - K_{PET}(= \kappa_{PET} \frac{A_{PET}}{l_{PET}}) \right)$, and the in-plane $\kappa$ was calculated from $K_{CNT}$ ($\kappa_{CNT} = K_{CNT} \frac{l_{PET}}{A_{PET}}$). A thin PET film with a thickness of 12 μm was used to ensure a sufficiently large $K_{CNT}/K_{CNT/PET}$ ratio and to maximize the measurement sensitivity. The dimensions of the samples such as the length and width were determined using an optical microscope, while the thicknesses of Films 1 and 2, which were on the order of 100 nm, were measured by atomic force microscopy.

The cross-plane $\kappa$ of Film 1 was measured using the time domain thermoreflectance (TDTR)[5,6], which is a well-established method that operates with pulsed laser and pump-and-probe techniques to characterize thermal transport of thin films and interfaces. The film was coated with a ~100-nm-thick Al transducer film. The details of our TDTR setup are described elsewhere[7].

The environment temperature was strictly controlled by Mercury iTC (Oxford Instruments) and the temperature fluctuation for each measurement was kept below 0.1 K to suppress deviation of the signal.

*Simulations:*

We used the atomic Green's function (AGF) method to calculate the thermal properties of a perfectly aligned CNT bundle both in the alignment direction and in the cross-sectional direction. Details of the AGF method are described in Reference [8]. As the average diameter of consisting CNT is 1.4 nm, a hexagonal unit cell consisting of (10,10) single-wall CNTs (diameter=1.36 nm) was used as a representative atomic scale model for the calculations. As shown in Figure S3a and S3b, the prepared cells for the alignment and perpendicular directions, respectively, were repeated twice between the lead sections. Periodic boundary conditions were applied in the directions of the cross section (Figure S3c and S3d). The interatomic interactions within and between CNTs were modeled by the Tersoff[9] and Lennard-Jones potentials[10], respectively; the potential parameters are described in the references. The transport calculations by the AGF method assumed fully ballistic phonon transport.

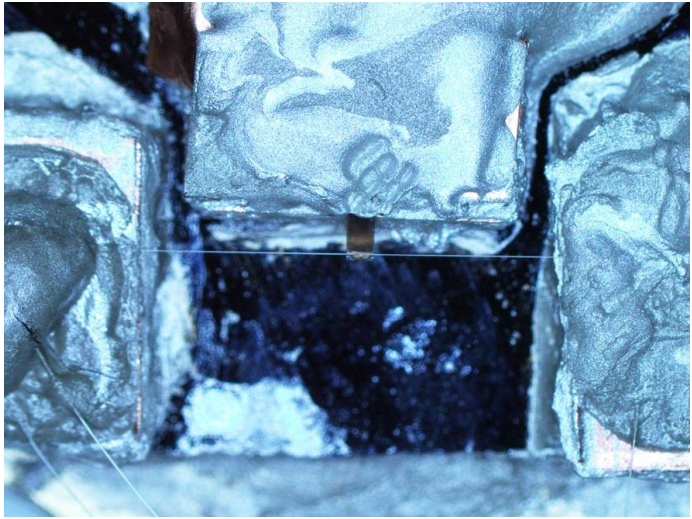

Figure S1. Experimental setup of T-type method.

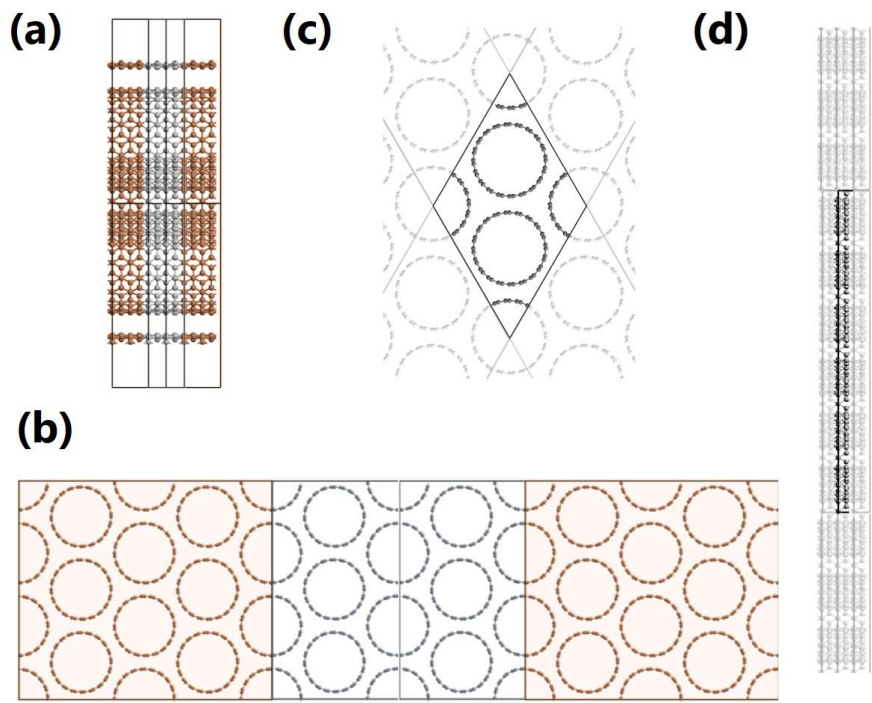

Figure S2. Atomic scale model of bundled (10, 10) single-walled CNTs for the atomic Green's function (AGF) method calculation in the (a) aligned and (b) perpendicular direction. The unit cells are sandwiched by colored lead section. (c,d) The boundary condition applied in the directions of the cross section.

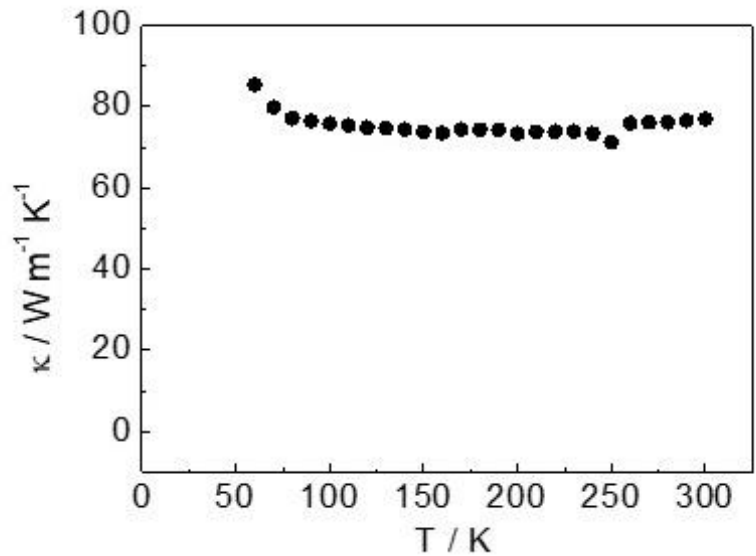

Figure S3. Temperature dependence of Pt measured by joule heating.

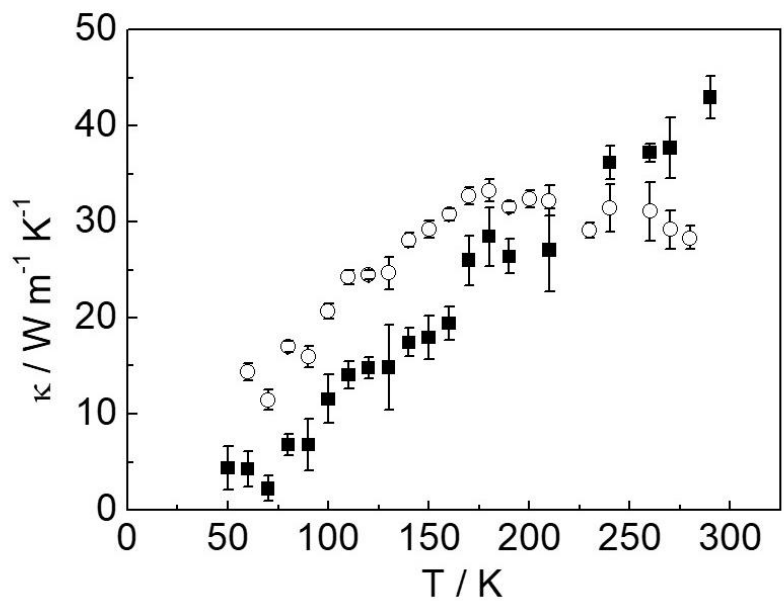

Figure S4. Measured temperature dependence of thermal conductivity. ■: Film 1 (alignment direction), ○: Film 2 (alignment direction).

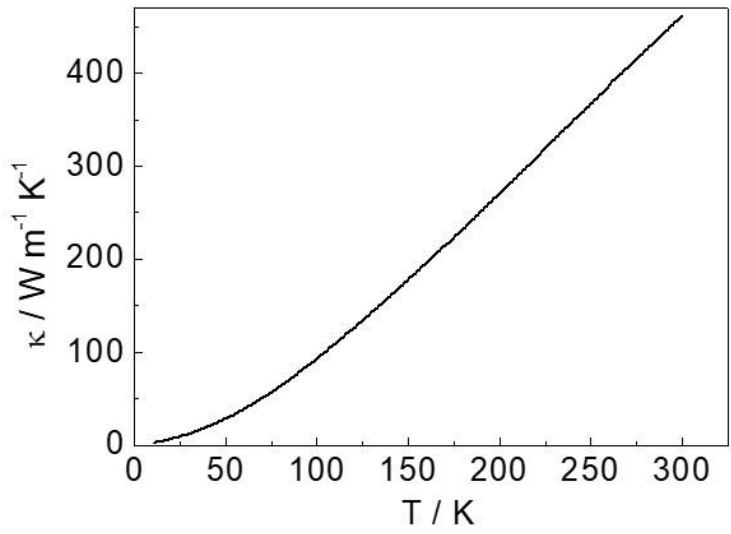

Figure S5. Temperature dependence of simulated effective thermal conductivity of Film 1.

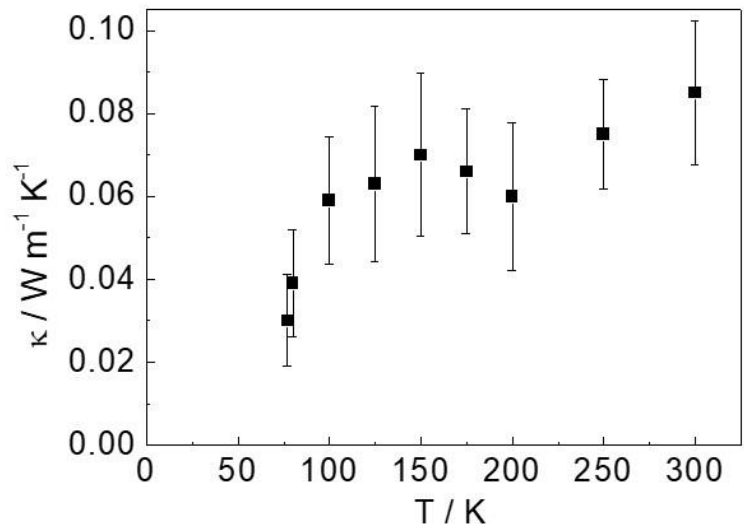

Figure S6. Measured thermal conductivity of Film 1 in the perpendicular direction in the temperature range of 77 to 300 K.

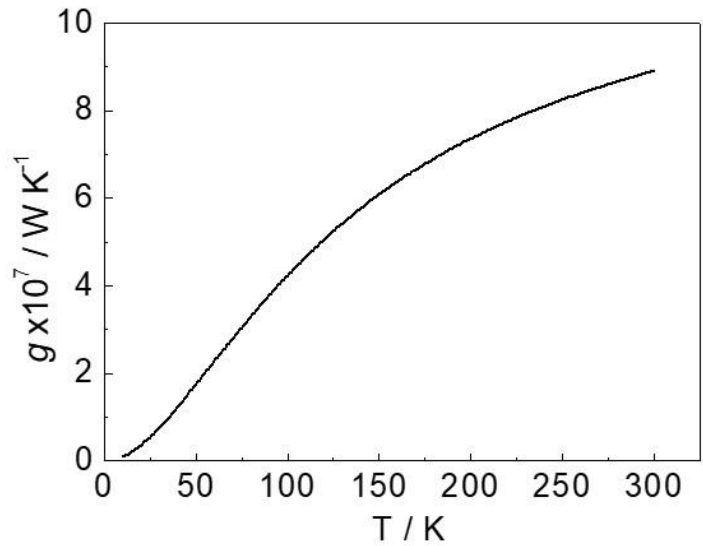

Figure S7. Simulated thermal conductance of Film 1 in the perpendicular direction in the temperature range of 10 to 300 K.